\documentclass[a4paper]{jpconf} %V4
\usepackage{graphicx}

\newcommand{\out}[1]{}

\newcommand{\vek}[1]{%
        \hbox{\textbf #1}}

\newcommand{\svek}[1]{%
        {\mathbf #1}}
\newcommand{\bra}[1]{\ensuremath{\langle #1|}}
\newcommand{\ket}[1]{\ensuremath{|#1\rangle}}
\newcommand{\braket}[2]{\ensuremath{\langle #1|#2\rangle}}

\newcommand{\pr}{%
        ^\prime}
\newcommand{\prpr}{%
        ^{\prime \prime}}

\begin{document}
\title{QS{\it GW}+DMFT: an electronic structure scheme for the iron pnictides and beyond}

\author{Jan M.\ Tomczak}

\address{Institute of Solid State Physics, Vienna University of Technology, A-1040 Vienna, Austria}

\ead{jan.tomczak@tuwien.ac.at}

\begin{abstract}
%lemme sell you this vacuum cleaner! It is AMAZING!
While in strongly correlated materials one often focuses on {\it local} electronic 
correlations, 
the influence of {\it non-local} exchange and correlation effects beyond band-theory can be
pertinent in systems with more extended
orbitals. Thus in many compounds an adequate theoretical description requires the joint treatment of 
local and non-local self-energies. Here, I will argue that this is the case for the iron pnictide and chalcogenide superconductors.
As an approach to tackle their electronic structure, 
I will detail the implementation of the recently proposed scheme that combines the quasi-particle self-consistent {\it GW} approach with dynamical mean-field theory:
QS{\it GW}+DMFT. I will showcase the possibilities of QS{\it GW}+DMFT with an application on BaFe$_2$As$_2$.
Further, I will discuss the empirical finding that in pnictides dynamical and non-local correlation effects separate
within the quasi-particle band-width.
\end{abstract}

\section{Introduction}

%``Most of the concepts of chemistry are local concepts''\cite{Anderson1984311}

Many phenomena in correlated materials --the Kondo effect, the Mott insulator, etc.-- can be described by {\it local} correlations.
Such is the source of success of dynamical mean field theory (DMFT)\cite{bible,RevModPhys.78.865} for the quintessential physics in
transition metals\cite{licht_katsnelson_kotliar}, their oxides\cite{PhysRevLett.86.5345,tomczak_v2o3_proc,me_psik,PhysRevLett.102.146402} and silicides\cite{jmt_fesi}, 
or rare-earth\cite{PhysRevLett.87.276404,PhysRevLett.102.096401,jmt_cesf} and actinide\cite{deltaPu,PhysRevB.84.195111} compounds. 

With experiments indicating sizable correlation effects in the newly discovered iron pnictides and chalcogenides\cite{RevModPhys.83.1589}, DMFT 
was swiftly called upon and delivered explanations for enhanced effective masses\cite{1367-2630-11-2-025021,PhysRevB.80.085101,PhysRevB.82.064504,Yin_pnictide,PhysRevB.85.094505,werner_bfa,annurev-conmatphys-020911-125045}, lower-than-expected ordered moments\cite{Yin_pnictide,PhysRevB.81.220506}, as well as previously elusive structural aspects\cite{PhysRevB.84.054529}.

However, for these systems both experimental and theoretical evidence has emerged
that {\it non-local} exchange and correlation effects are non-negligible, and at times even pivotal for a qualitative understanding.
I will name  
two specific examples: (1) The Fermi surface of LiFeAs has been reported to have a distinctly different topology (number of sheets) in experiment\cite{PhysRevLett.105.067002} than in 
density functional theory (DFT) or DMFT\cite{Yin_pnictide,PhysRevB.85.094505}. 
(2) The size of electron and hole pockets in BaFe$_2$As$_2$ are largely overestimated in DFT and DMFT
\cite{Yin_pnictide,PhysRevLett.110.167002}.
Thus quantities that are sensitive to excitations near the Fermi level, such as the thermopower, cannot be reproduced.

\begin{figure}[h]
\includegraphics[width=14pc]{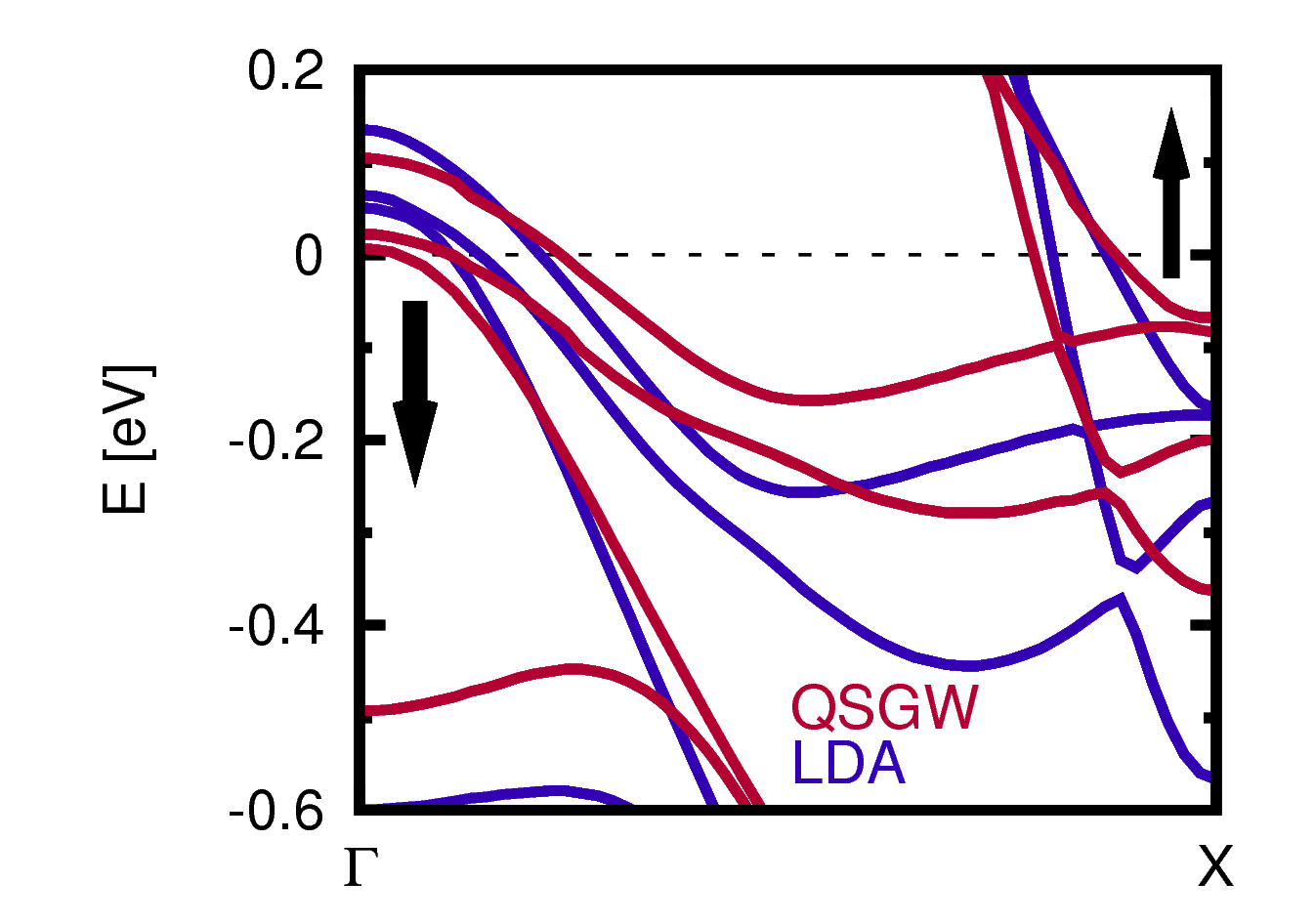}\hspace{2pc}%
\begin{minipage}[b]{18.5pc}\caption{\label{bfa1}
BaFe$_2$As$_2$: Comparison of the band-structures of DFT(LDA) with QS{\it GW}.
Arrows indicate the shrinking of both the hole and electron pockets at $\Gamma$ and $X$, respectively, in congruence with experiment (see e.g.\ Refs.~\cite{PhysRevB.83.054510,PhysRevLett.110.167002})
and not captured within DMFT.}
\end{minipage}
\end{figure}

On the other hand, I have recently shown\cite{jmt_pnict} that using quasi-particle self-consistent (QS) {\it GW} \cite{PhysRevLett.93.126406,schilfgaarde:226402}, a many-body perturbation theory\cite{RevModPhys.74.601} that excels at treating non-local exchange and also includes some non-local correlation effects, these problems are solved: (1) For LiFeAs QS{\it GW} yields a Fermi surface in excellent in agreement with experiment,
(2) a non-local self-energy shift shrinks both electron and hole pockets in BaFe$_2$As$_2$ to about half their size with respect to DFT, as shown in \fref{bfa1}.

Many-body perturbation theory in its QS{\it GW} realization thus pinpoints non-local exchange and correlation
effects to be crucial for the description of many aspects in the iron pnictides and chalcogenides.
Nevertheless there are several properties that require the capturing of local dynamical renormalizations 
beyond the possibilities of a perturbative treatment\cite{jmt_pnict}: Indeed quasi-particle weights come out too big in {\it GW},
electron lifetimes are overestimated, and incoherence effects at finite temperatures are out of reach.

For these reasons, combining the best of both methodologies, {\it GW} and DMFT, is a promising route in electronic structure theory\cite{PhysRevLett.90.086402,PhysRevLett.92.196402}.
Owing to the complexity of the {\it GW}+DMFT approach\cite{PhysRevLett.90.086402}, however, applications are hitherto scarce:
Fully dynamical and self-consistent implementations have been achieved for one-band systems only\cite{ayral_gwdmft,PhysRevB.87.125149,PhysRevLett.110.166401}.
For other, realistic materials calculations, only lately an implementation that includes the pivotal dynamics of the Hubbard interaction has been pioneered and
applied to SrVO$_3$\cite{jmt_svo,jmt_svo_extended}\footnote{For calculations for SrVO$_3$ with static interactions and additional approximations, see Ref.~\cite{PhysRevB.88.165119}; and Ref.~\cite{PhysRevB.88.235110} for an {\it ad hoc} combination of {\it GW} and DMFT self-energies.}.

Recently, we proposed to merge QS{\it GW} with DMFT in a ``QS{\it GW}+DMFT'' approach\cite{jmt_pnict} that retains most of the advantages of {\it GW}+DMFT\cite{PhysRevLett.90.086402}
at a significantly reduced computational complexity.
Here, after briefly reviewing QS{\it GW}, I will give practical details on how to implement the QS{\it GW}+DMFT scheme and showcase its potential for the case of BaFe$_2$As$_2$.
Further, I will discuss the recent finding that in many systems, non-local and dynamical correlations can be separated for all practical purposes. This non-trivial result
paves the way for the devising of physically motivated approximative electronic structure scheme, such as DMFT@non-local-{\it GW}\cite{jmt_svo_extended}
and SEX+DMFT\cite{paris_sex} that I will comment on briefly.

\section{QS{\it GW}+DMFT -- the formalism}

\subsection{QS{\it GW}: a self-consistent many-body perturbation theory}

The central quantity for spectral properties of correlated materials is the one-particle Greens-function $G$, which can be expressed as
\begin{equation}
G^{-1}(\svek{k},\omega) =\omega-H^0(\svek{k})-\Sigma(\svek{k},\omega)
\label{G}
\end{equation}
where we assume a matrix structure in the space of orbitals.
Here, $H^0$ is a reference effective one-particle Hamiltonian, and $\Sigma$ a momentum and frequency dependent self-energy 
that is defined with respect to exchange and correlation effects already included in $H^0$.

For example, 
in DFT+DMFT\cite{RevModPhys.78.865}, 
$H^0=H^{\hbox{\tiny DFT}}=-\nabla^2+v_{crystal}+v_{Hartree}+v_{xc}^{\hbox{\tiny DFT}}$ 
is the Kohn-Sham Hamiltonian, and 
$\Sigma^{\hbox{\tiny DFT+DMFT}}(\omega)=\left[\Sigma^{\hbox{\tiny DMFT}}(\omega)-E_{dc}\right]_{LL\pr RR}\ket{RL}\bra{R L\pr}$
is the
DMFT self-energy that is local (momentum-independent) when expressed in a local basis set of correlated orbitals $\braket{\svek{r}}{RL}=\chi_{RL}(\svek{r})$,
labelled by a lattice site or unit-cell $R$ and an orbital index $L$.
From $\Sigma^{\hbox{\tiny DMFT}}$ the so-called ``double counting'' correction, $E_{dc}$, that mimics self-energy effects already contained in $H^0$ must be
subtracted. Several proposals and rules of thumb for the choice of $E_{dc}$ exist. However, its exact form is unknown,
introducing some arbitrariness to DFT+DMFT.\\
Within the QS{\it GW} approach\cite{PhysRevLett.93.126406,schilfgaarde:226402}, 
\begin{equation}
H^0=H^{\hbox{\tiny QS{\it GW}}}=-\nabla^2+v_{crystal}+v_{Hartree}+v_{xc}^{\hbox{\tiny QS{\it GW}}}.
\label{HQSGW}
\end{equation}
The {\it GW} self-energy, $\Sigma^{\hbox{\tiny \it GW}}=G^{\hbox{\tiny QS{\it GW}}}_0\cdot W$,  
where ${G^{\hbox{\tiny QS{\it GW}}}_0}^{-1}=  \omega-H^{\hbox{\tiny QS{\it GW}}}$,
enters with respect to the static, yet non-local and orbital-dependent QS{\it GW} exchange-correlation potential, $v_{xc}^{\hbox{\tiny QS{\it GW}}}$:
\begin{equation}
\Sigma^{\hbox{\tiny QS{\it GW}}}=\Sigma^{\hbox{\tiny \it GW}}-v_{xc}^{\hbox{\tiny QS{\it GW}}} 
\label{SigGW}
\end{equation}
The potential $v_{xc}^{\hbox{\tiny QS{\it GW}}}$
is determined by the requirement that 
the quasi-particle energies $E_{\svek{k}j}$, given by the poles of the interacting Greens-function \eref{G},
\begin{equation}
\det\left[E_{\svek{k}j}-H^{\hbox{\tiny QS{\it GW}}}(\vek{k})-\Re\Sigma^{\hbox{\tiny QS{\it GW}}}(\vek{k},E_{\svek{k}j})\right]\stackrel{!}{=}0
\label{QPE}
\end{equation}
coincide with 
the eigenvalues of $H^{\hbox{\tiny QS{\it GW}}}$, i.e.\ 
$H^{\hbox{\tiny QS{\it GW}}}(\svek{k})\ket{\Psi_{\svek{k}j}}=E_{\svek{k}j}\ket{\Psi_{\svek{k}j}}$.
This means that at self-consistency
$H^{\hbox{\tiny QS{\it GW}}}(\svek{k})+\Re\Sigma^{\hbox{\tiny QS{\it GW}}}(\svek{k},E_{\svek{k}j})=H^{\hbox{\tiny QS{\it GW}}}(\svek{k})$, 
or, equivalently,
%
%\begin{equation}
$\bra{\Psi_{\svek{k}j}}\Sigma^{\hbox{\tiny \it GW}}(\svek{k},E_{\svek{k}j})-v_{xc}^{\hbox{\tiny QS{\it GW}}}\ket{\Psi_{\svek{k}j}}=0$.
%\label{eq:}
%\end{equation}
%
The effective QS{\it GW} potential thus in particular incorporates dynamical renormalizations of the band dispersions (e.g.\ the $Z$-factor in the Fermi liquid regime)
through a state and momentum dependent static shift.
It can be shown that %quasi-particle 
self-consistency can be reached iteratively by 
hermitianizing
the self-energy\cite{PhysRevB.76.165106}: 
\begin{equation}
v_{xc}^{\hbox{\tiny QS{\it GW}}}=\frac{1}{2}\sum_{ij\svek{k}}\ket{\Psi_{\svek{k}i}}\Re\left[\Sigma^{\hbox{\tiny \it GW}}_{ij}(\svek{k},E_{\svek{k}i})+\Sigma^{\hbox{\tiny \it GW}}_{ji}(\svek{k},E_{\svek{k}j})  \right]\bra{\Psi_{\svek{k}j}}.
\label{qsGW}
\end{equation}
The virtue of QS{\it GW} is that in the end all remnants of the DFT starting point are gone, and the quasi-particle energies are unambiguously defined.
Moreover, it was recently demonstrated that the above self-consistency procedure --irrespective of whether it is performed within 
the {\it GW} approximation
or any other many-body method-- converges to the best possible effective one-particle theory accessible to the employed many-body technique\cite{2014arXiv1406.0772I}.%
\footnote{%
While the effective one-particle Hamiltonian, $H^{\hbox{\tiny QS{\it GW}}}$, neglects life-time effects, %as encoded in the imaginary part of the self-energy,
$\Im\Sigma^{\hbox{\tiny QS{\it GW}}}$ is included when computing spectra via the interacting Greens-function \eref{G}: 
${G^{\hbox{\tiny QS{\it GW}}}}^{-1}(\svek{k},\omega)$$=$$\omega-H^{\hbox{\tiny QS{\it GW}}}(\svek{k})$$-$$\Sigma^{\hbox{\tiny QS{\it GW}}}(\svek{k},\omega)$.
}

\subsection{QS{\it GW}+DMFT}

The general strategy in QS{\it GW}+DMFT is to take out of the improved effective one-particle Hamiltonian, $H^{\hbox{\tiny QS{\it GW}}}$,
all renormalizations that are propelled by {\it local} correlations.
In the spirit of the above discussion, one is tempted to use
$H^0=H^{\hbox{\tiny QS{\it GW}}}$ 
and define the self-energy in a local basis $\left\{\chi_{RL}(\svek{r})\right\}$ as
%
%\begin{equation}
$\Sigma^{\hbox{\tiny QS{\it GW}+DMFT}}(\omega)=\left[\Sigma^{\hbox{\tiny DMFT}}(\omega)-\Re\Sigma^{\hbox{\tiny {\it GW}}}_{local}(\omega)\right]_{LL\pr} \ket{RL}\bra{RL\pr}$,
with the local projection $\Sigma_{LL\pr\, local}^{\hbox{\tiny {\it GW}}}(\omega)=\sum_{\svek{k}}\Sigma_{LL\pr}^{\hbox{\tiny {\it GW}}}(\svek{k},\omega)$.
%\label{qsgwdmft2}
%\end{equation}
%
Indeed the quasi-particle Hamiltonion, $H^{\hbox{\tiny QS{\it GW}}}$, 
includes via \eref{qsGW} all {\it hermitian} self-energy effects, 
the local parts of which have to be taken out.
However, in this naive ansatz the total self-energy $\Sigma^{\hbox{\tiny DMFT}}-\Re\Sigma^{\hbox{\tiny {\it GW}}}_{local}$ 
manifestly violates the Kramers-Kronig relations, i.e.\ causality.

Here, we propose two natural schemes to construct a quasi-particle Hamiltonian, $H^{\hbox{\tiny QS{\it GW}}}_{non-local}$, that is exactly double-counting free when combined with a local (DMFT) self-energy.
As a first method, I shall detail an intuitive scheme that requires the hermitian part of the {\it GW} self-energy to be linear in frequency over the
quasi-particle band-width. In a second step, a general framework will be introduced.

\subsubsection{unrenormalize $H^{QSGW}$ to linear order in the local self-energy.}

Suppose we already dispose of the Hamiltonian $H_{non-local}^{\hbox{\tiny QS{\it GW}}}$ in which only non-local correlations effects have been accounted for.
Within QS{\it GW}, the latter is further renormalized by local correlations described by $\Sigma_{local}^{\hbox{\tiny {\it GW}}}(\omega)$.
The interacting Greens-function then can be expressed with these two quantities:
\begin{eqnarray}
{G^{\hbox{\tiny QS{\it GW}}}}^{-1}(\svek{k},\omega)=\omega-H_{non-local}^{\hbox{\tiny QS{\it GW}}}(\svek{k}) -\Sigma_{local}^{\hbox{\tiny {\it GW}}}(\omega)
\label{pole}
\end{eqnarray}
Performing a low-energy expansion, $\Sigma_{local}^{\hbox{\tiny {\it GW}}}(\omega)$$\approx$$\Re\Sigma_{local}^{\hbox{\tiny {\it GW}}}(\omega$$=$$0)$$+$$ (1-Z_{loc}^{-1}) \omega$,
in the local basis, and assuming this Fermi liquid-like self-energy to extend over the range of the quasi-particle dispersion, \eref{pole} becomes 
%
%\begin{eqnarray}
$\omega-Z_{loc}\left[H_{non-local}^{\hbox{\tiny QS{\it GW}}}(\svek{k}) +\Re\Sigma_{local}^{\hbox{\tiny {\it GW}}}(0)\right]$
%\label{pole2}
%\end{eqnarray}
%
from which we identify the full %QS{\it GW} 
Hamiltonian
%
%\begin{eqnarray}
$H_{\phantom{bla}}^{\hbox{\tiny QS{\it GW}}}(\svek{k}) =  Z_{loc} \left [ H_{non-local}^{\hbox{\tiny QS{\it GW}}}(\svek{k}) + \Re\Sigma_{local}^{\hbox{\tiny {\it GW}}}(0)  \right]$,
%\label{pole3}
%\end{eqnarray}
%
whence
\begin{equation}
H_{non-local}^{\hbox{\tiny QS{\it GW}}}(\svek{k})  = %\left( 
Z_{loc}^{-1} \,
%\right)%_{LL\prpr} 
H_{\phantom{L\prpr L\pr}}^{\hbox{\tiny QS{\it GW}}}(\svek{k}) -\Re\Sigma_{local}^{\hbox{\tiny {\it GW}}}(\omega=0) 
\label{unrenorm}
\end{equation}
This expression is very intuitive: The quasi-particle bandwidth, that has been narrowed by dynamical renormalizations, is {\it widened} by the inverse of the local component of the quasi-particle weight $Z_{loc}$.
Additionally, local correlation-induced crystal-field splittings are subtracted.
The local self-energy that will be provided by the DMFT will go these steps backwards and add the more reliable local shifts and quasi-particle renormalization
factor $Z^{\hbox{\tiny DMFT}}$.
While in many systems the hermitian part of the {\it GW} self-energy will be linear over the band-width of low-energy orbitals,
making the just described {\it linear} QS{\it GW}+DMFT prescription reliable for all practical purposes, a more general procedure
is possible and will be described in the following.

\subsubsection{non-local quasi-particlization.}
Again, instead of devising a separate double counting $E_{dc}$ {\`a} la DFT+DMFT, the idea is to construct a quasi-particle Hamiltonian, $H_{non-local}^{\hbox{\tiny QS{\it GW}}}$, that includes all non-local exchange and correlation effects beyond DFT,
but is void of local correlations. 
In a local basis, $\{\chi_{RL}(\svek{r})\}$, we first determine local and non-local projections of the self-energy, 
$\Sigma_{local}^{\hbox{\tiny {\it GW}}}=\sum_{\svek{k}}\Sigma^{\hbox{\tiny {\it GW}}}$, $\Sigma_{non-local}^{\hbox{\tiny {\it GW}}}=\Sigma^{\hbox{\tiny {\it GW}}}-\Sigma_{local}^{\hbox{\tiny {\it GW}}}$,
respectively. Then we transform these back into the band representation $\{\Psi_i(\svek{r})\}$, use \eref{SigGW}, and
identify the respective terms in %the quasi-particle 
equation \eref{QPE}:
\begin{equation}
\det\left[ \omega-H^{\hbox{\tiny QS{\it GW}}}(\svek{k})-\Re\Sigma_{non-local}^{\hbox{\tiny {\it GW}}}(\svek{k},\omega)-\Re\Sigma_{local}^{\hbox{\tiny {\it GW}}}(\omega)+v_{xc}^{\hbox{\tiny QS{\it GW}}}(\svek{k}) \right]=0
\label{v1}
\end{equation}
Now we want to replace $\Re\Sigma_{ij\, non-local}^{\hbox{\tiny {\it GW}}}(\svek{k},\omega)$ by a {\it static} term
that does not modify the energies $E_{ki}$ that are the solutions of \eref{v1}. This is exactly the spirit the quasi-particle self-consistency\cite{PhysRevLett.93.126406,schilfgaarde:226402}, but now we construct a potential that 
only includes non-local parts of the self-energy. Akin to \eref{qsGW} the natural choice is:%
\footnote{We note that in the usual QS{\it GW} prescription, the operator $v_{xc}^{\hbox{\tiny QS{\it GW}}}$ of \eref{qsGW} does not always reproduce the quasi-particle energies exactly.
This is owing to band-off-diagonal  ($i\ne j$) elements of the self-energy. The resulting ambiguity in the self-consistent Hamiltonian was shown to be small\cite{PhysRevB.76.165106,PhysRevB.78.075106}.
In particular, once self-consistency is achieved, $\Sigma^{\hbox{\tiny {\it GW}}}(\svek{k},E_{\svek{k}i})$ should be diagonal in the band basis, and thus \eref{qsGW2} reliable and unambiguous.}%
\begin{equation} 
\widetilde{v}_{xc\, non-local}^{\hbox{\tiny QS{\it GW}}}=\frac{1}{2}\sum_{ij\svek{k}}\ket{\Psi_{\svek{k}i}}\Re\left[\Sigma_{ij\, non-loc}^{\hbox{\tiny {\it GW}}}(\svek{k},E_{\svek{k}i})+\Sigma_{ji\, non-loc}^{\hbox{\tiny {\it GW}}}(\svek{k},E_{\svek{k}j})  \right]\bra{\Psi_{\svek{k}j}}.
\label{qsGW2}
\end{equation}
Transforming \eref{qsGW2} into the local basis, $\{\chi_{RL}(\svek{r}) \}$, the Hamiltonian for direct use in DMFT is:
\begin{equation}
H_{non-local}^{\hbox{\tiny QS{\it GW}}}(\svek{k}) =H_{}^{\hbox{\tiny QS{\it GW}}}(\svek{k})+\widetilde{v}_{xc\, non-local}^{\hbox{\tiny QS{\it GW}}}(\svek{k})-v_{xc}^{\hbox{\tiny QS{\it GW}}}(\svek{k})
\label{Hnl}
\end{equation}
We stress that $\widetilde{v}_{xc\, non-local}^{\hbox{\tiny QS{\it GW}}}$ is {\it not} the local projection of $v_{xc}^{\hbox{\tiny QS{\it GW}}}$, 
as the non-local projection of the self-energy was done at arbitrary frequency $\omega$, {\it before} evaluating it at the quasi-particle energies in \eref{qsGW2}. See Ref.~\cite{jmt_svo_extended} for a discussion of this intricacy.
The term that takes out {\it local} correlation effects is therewith actually momentum-dependent --a clear necessity for 
$H_{non-local}^{\hbox{\tiny QS{\it GW}}}$ to have a wider band-width than $H_{}^{\hbox{\tiny QS{\it GW}}}$.

\begin{figure}[h]
\includegraphics[width=14pc]{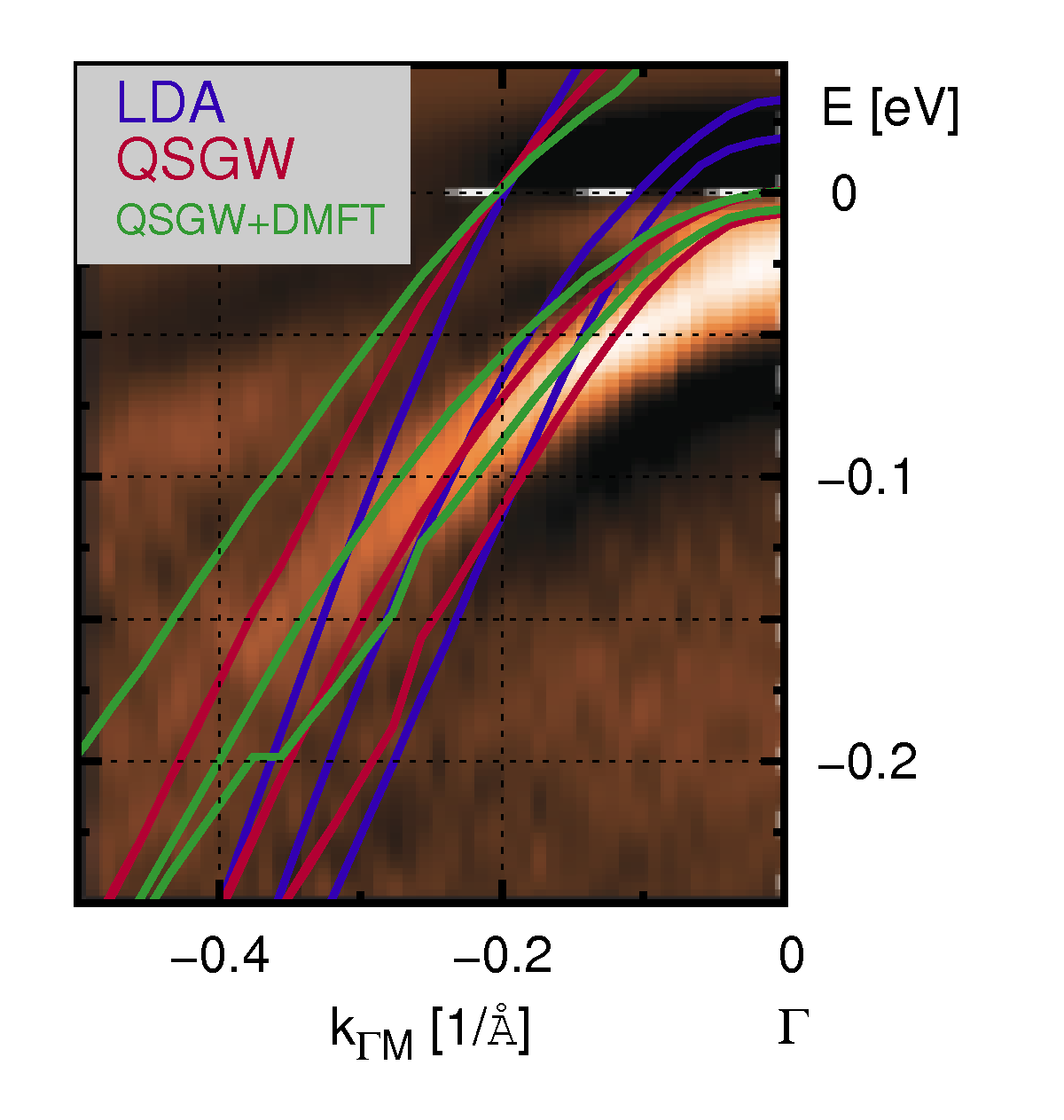}\hspace{2pc}%
\begin{minipage}[b]{16pc}\caption{\label{bfa2}
BaFe$_{1.85}$Co$_{0.15}$As$_2$: Comparison of photoemission\cite{PhysRevB.83.054510}, DFT(LDA), QS{\it GW}\cite{jmt_pnict} and 
a much simplified realization of QS{\it GW}+DMFT (see text for details).
Note that QS{\it GW}+DMFT describes both the size of the hole pockets and the dispersion (effective mass renormalization).
Electron doping was simulated in the virtual crystal approximation.\\[0.75cm]}
\end{minipage}
\vspace{-0.5cm}
\end{figure}

\section{Application}

In \fref{bfa2}, I show a simplified realization of QS{\it GW}+DMFT for electron-doped BaFe$_2$As$_2$.
In comparison with angle resolved photoemission (ARPES) experiments\cite{PhysRevB.83.054510} there are two major discrepancies in the DFT(LDA) band-structure:
(a) the size of hole pockets at the $\Gamma$ point is drastically overestimated, and (b) the dispersion is too pronounced, i.e.\ effective masses are too small.
DMFT does not correct problem (a) \cite{PhysRevLett.110.167002}, but it yields larger effective masses\cite{Yin_pnictide}.
QS{\it GW} on the other hand, substantially improves on (b) the size of the hole pockets, which is interpreted as an effect of a non-local self-energy.
The dispersion %in QS{\it GW} 
is however still too big with respect to experiment, owing to the perturbative treatment of the quasi-particle weight $Z$\cite{jmt_pnict}.

Mimicking QS{\it GW}+DMFT by substituting the local dynamical renormalizations (provided by the local quasi-particle weight $Z_{loc}$) in QS{\it GW}\cite{jmt_pnict} with published values from DMFT\cite{Yin_pnictide}, we see that
QS{\it GW}+DMFT combines {\it the best of both worlds}:
a good description of both, (a) the size of the pockets, and (b) the effective masses that renormalize the slope of the dispersion.
We note that the dispersion of the non-local QS{\it GW}, $H_{non-local}^{\hbox{\tiny QS{\it GW}}}$, is {\it larger} than in DFT(LDA), as the non-local self-energy delocalizes charge carriers.
The effective mass of QS{\it GW}+DMFT is thus lower than in DMFT\cite{jmt_pnict}. 
This will have a notable effect also on the spin degrees of freedom: 
Besides the correction of the energy scale associated with the size of the charge pockets\cite{Sales2010304}, the carrier delocalization will push the onset of single-ion physics
(Curie-like susceptibility) to higher temperatures. Indeed the magnetic susceptibility in the iron pnictides has a universal linear dependence
up to high temperatures\cite{RevModPhys.83.1589}.

Let me note in passing that also a second deficiency of DFT+DMFT methods is improved upon by QS{\it GW}+DMFT:
Indeed DMFT can only be applied to a low-energy subspace of orbitals (typically $d$ or $f$-electron states).
However, it is known that ligand states (e.g.\ the 2$p$-orbitals in transition metal oxides) are not well captured within DFT,
owing to inter- and out-of-subspace exchange and correlation effects\cite{ferdi_down,PhysRevB.87.195144}.
As an all-electron method, QS{\it GW} includes these effects; for instance it corrects the position of the Se-4$p$ states in the chalcogenide FeSe 
by almost 1~eV\cite{jmt_pnict} in excellent agreement with experiment\cite{PhysRevB.82.184511}.

\section{Relation of QS{\it GW}+DMFT to full {\it GW}+DMFT}

The QS{\it GW}+DMFT approach, proposed in Ref.~\cite{jmt_pnict} and detailed above, offers to retain salient features of the full {\it GW}+DMFT methodology\cite{PhysRevLett.90.086402}, 
and to remedy some of the pathologies of current {\it GW}+DMFT implementations: 
State-of-the-art {\it GW} codes  rely on a non-interacting form of the Greens-function,
complicating the feedback of the DMFT self-energy and susceptibility onto the {\it GW} part.
Therefore, hitherto this has been achieved only in the one band case\cite{ayral_gwdmft,PhysRevB.87.125149,PhysRevLett.110.166401}.
Indeed all other, realistic {\it GW}+DMFT calculations so far\cite{PhysRevLett.90.086402,jmt_svo,jmt_svo_extended} fix non-local correlations to the one-shot {\it GW} level
 and the screened interaction to constraint random phase approximation (cRPA)\cite{PhysRevB.70.195104} results. These quantities thus depend on the effective one-particle Greens-function chosen as the starting point, most often provided by DFT.
QS{\it GW}+DMFT circumvents global iterations by limiting the self-consistency of non-local correlation effects to the QS{\it GW} level%
\footnote{%
Although it is absolutely conceivable to include the hermitian part of the DMFT self-energy in the construction of the effective exchange and correlation potential \eref{qsGW}.
}%
.
The pathology of a DFT starting point dependence
is thus avoided%
\footnote{%
Also, there is a limited (one-particle) feedback of local correlations onto the screening of the Hubbard interaction. 
}.

In (QS){\it GW}+DMFT the double counting, $E_{dc}$, is known exactly.
Its static form in the QS{\it GW}+DMFT approach allows for a  Hamiltonian description.
Indeed, $H_{non-local}^{\hbox{\tiny QS{\it GW}}}$ of equation \eref{Hnl} or \eref{unrenorm}, can be viewed as a vastly improved effective one-particle starting point\cite{2014arXiv1406.0772I} for DMFT.
Consequently, current DFT+DMFT implementations can be used with little or no  modification. 
This also avoids handling the momentum and frequency dependent {\it GW} self-energy in the DMFT self-consistency, 
drastically reducing memory requirements.

The short-cuts introduced by QS{\it GW}+DMFT come with the cost
of neglecting 
(a) the momentum variation of quasi-particle coherence%
\footnote{that is negligible on the {\it GW} level, and is found to be small 
in 3 dimensions even when using non-perturbative techniques\cite{jmt_dga3d}. This is because most momentum-dependent correlations are static\cite{jmt_dga3d}.
Moreover, restriction (a) can be lifted in a more sophisticated implementation of QS{\it GW}+DMFT: In the spirit of \eref{QPE},
 the non-local projection of the fully non-hermitian and frequency-dependent {\it GW} self-energy, $\Sigma^{\hbox{\tiny QS{\it GW}}}_{non-loc}$, does not change the quasi-particle energies
of \eref{Hnl}. Hence $\Sigma^{\hbox{\tiny QS{\it GW}}}_{non-loc}$ can simply be added on top of \eref{Hnl} in the DMFT self-consistency.
}%
, %and, in case a fully self-consistent {\it GW}+DMFT implementation is achieved one day,
(b) local vertex corrections in the polarization beyond RPA%
\footnote{%
these local vertex corrections have been found to be important in a study of the extended Hubbard model\cite{ayral_gwdmft,PhysRevB.87.125149}.
}%
, and
(c) non-perturbative corrections to the {\it GW} one-particle propagator%
\footnote{%
using interacting Greens-functions in self-consistent {\it GW} calculations has been found to be detrimental for spectral properties\cite{PhysRevB.58.12684}.
It can be expected that adding local vertex corrections via DMFT improves this. % situation.
}%
.

\begin{figure}[h]
\includegraphics[width=16pc,angle=0]{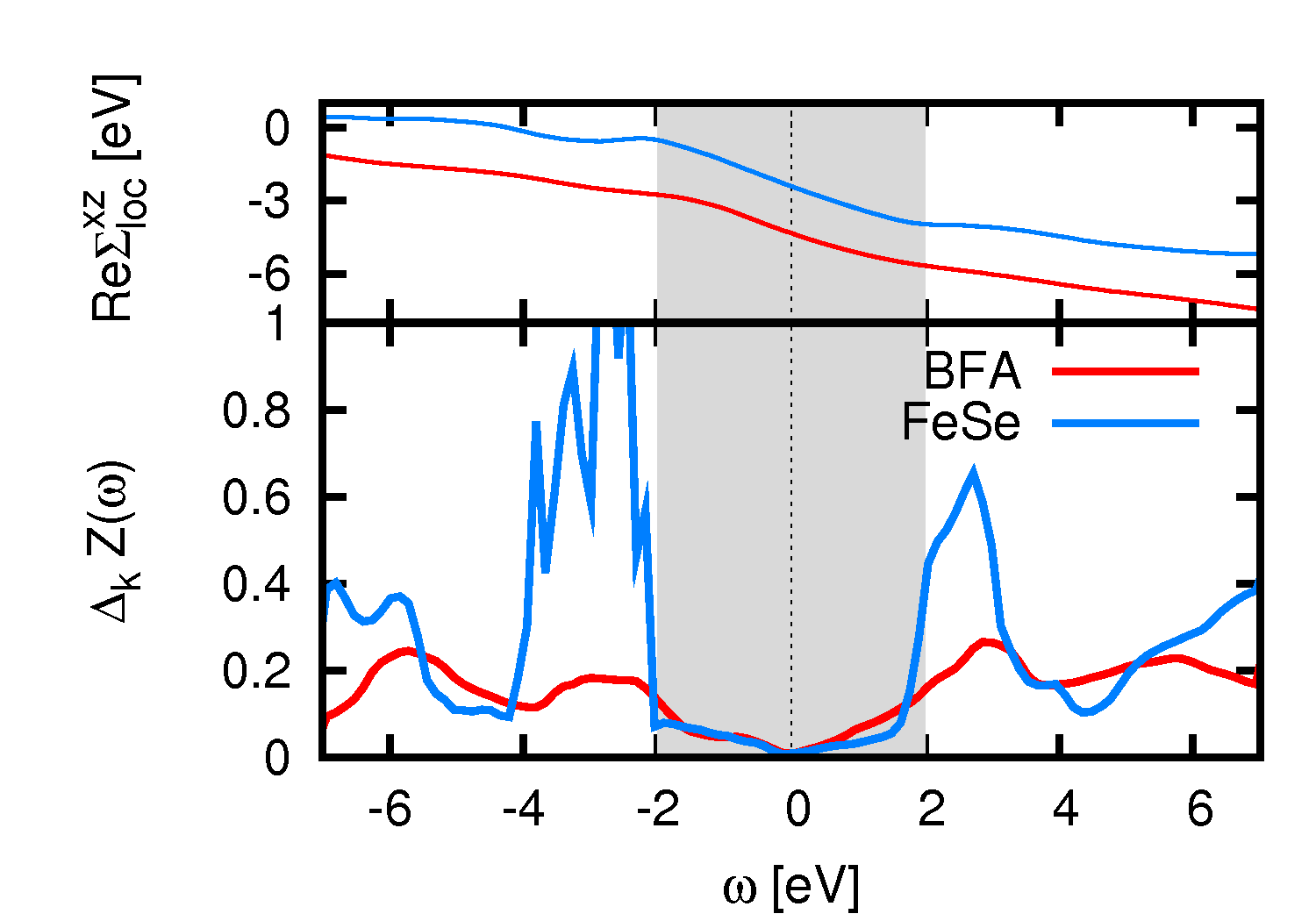}\hspace{2pc}%
\begin{minipage}[b]{19pc}\caption{\label{Zw}
BaFe$_2$As$_2$ and FeSe: Real parts of the local QS{\it GW} self-energy ($xz$ Wannier component, upper panel)
and the standard deviation $\Delta_kZ$ of the generalized quasi-particle weight $Z^{\svek{k}}(\omega)$$=$$[1-\partial_\omega\Re\Sigma(\svek{k},\omega)]^{-1}$ with respect to momentum
(lower panel): The k-dependence of the dynamical self-energy is negligible in the (gray shaded) regime in which $\Re\Sigma$ is linear.
}
\end{minipage}
\end{figure}

\section{Separability of non-local and dynamical correlations}

When not using the self-consistency prescription \eref{qsGW},
for a Hamiltonian formalism to include non-local correlation effects necessitates that correlations are {\it separable}
into non-local and dynamical contributions\cite{jmt_svo_extended}.
Indeed, a self-energy %defined with respect to a Hamiltonian $H^0$ 
can formally be split into
$\Sigma^x(\svek{k})+\Sigma^c_{non-loc}(\svek{k},\omega)+\Sigma^c_{loc}(\omega)$,
i.e.\ an exchange (``x'') and a correlation (``c'') part, in the latter of which we can isolate local (``loc'') and non-local (``non-loc'') components.
Only if the non-local part of the correlation self-energy is static, i.e.\ 
\begin{equation}
\Sigma^c(\svek{k},\omega)=\Re\Sigma^c_{non-loc}(\svek{k})+\Sigma^c_{loc}(\omega),
\label{separate}
\end{equation}
can we construct $H^{qp}_{non-loc}(\svek{k})=H^0(\svek{k})+\Sigma^x(\svek{k})+\Re\Sigma^c_{non-loc}(\svek{k})$ for use in DMFT.
Requirement \eref{separate} is a non-trivial statement: it is known to fail in 1D \& 2D systems.
Recently, however, it was shown that correlation effects in 3D systems such as the iron pnictides and chalcogenides\cite{jmt_pnict} and metallic transition metal oxides\cite{jmt_svo_extended}
verify property \eref{separate} within {\it GW} for energies inside the $d$-electron bandwidth.
This is exemplified in \fref{Zw} for BaFe$_2$As$_2$ and FeSe.
There, as a measure for the non-locality of dynamical renormalizations we plot the standard deviation
$\Delta_kZ(\omega)=\sqrt{\sum_{kL} \left| Z^{\svek{k}}_{LL}(\omega)-Z^{loc}_{LL}(\omega)\right|^2}$ 
of the momentum resolved quasi-particle weight $Z_{\svek{k}}$ with respect to its local value $Z_{loc}$, both of which we 
formally extend to finite frequencies: $Z_{\svek{k}}(\omega)=[1-\partial_\omega\Re\Sigma(\svek{k},\omega)]^{-1}$.
Indeed the momentum dependence virtually vanishes at the Fermi level, $\Delta_kZ(\omega=0)\approx 0$, and remains very low in the regime in which the
hermitian parts of the self-energy are linear.
This shows that in this energy range, non-local correlations are static, and dynamical correlations thus local%
\footnote{%
We note that this is true beyond the perturbative {\it GW} methodology, as was shown for the 3D Hubbard model\cite{jmt_dga3d}.
}%
.

This separability has two far-reaching consequences:
(a) it provides a highly encouraging {\it a posterori} justification for DMFT-based approaches that describe correlation effects with a dynamical, yet local, self-energy.
(b) Identifying that the sizable non-local correlations are dominantly static in nature may pave the way for physically motivated approximate electronic structure schemes,
such as the recently proposed DMFT@non-local-{\it GW}\cite{jmt_svo_extended}
 and SEX+DMFT\cite{paris_sex}.
In DMFT@nonlocal {\it GW}, the separability \eref{separate} is used to construct a non-locally renormalized Hamiltonian $H^{qp}_{non-loc}(\svek{k})$ 
from a one-shot {\it GW} calculation. It was shown that this approach successfully reproduces for SrVO$_3$ the quasi-particle dispersion of full {\it GW}+DMFT calculations,
and, if a dynamic Hubbard interaction is used, even the position of satellite features such as the plasmons peak is captured\cite{jmt_svo_extended}.
SEX+DMFT\cite{paris_sex}  identifies exchange effects beyond DFT as crucial non-local self-energy contributions and computes
a screened exchange (SEX) ingredient to supplement the one-particle Hamiltonian. Therewith \eref{separate} is fulfilled by construction.
The approach has been shown to be very successful in the description of ARPES experiments for BaCo$_2$As$2$\cite{paris_sex}.

Besides the exchange part, there potentially are, in correlated systems, momentum-dependent {\it correlation} effects%
\footnote{%
Indeed it was shown for the electron gas that corrections to the local 
approximation of the Coulomb hole are significant 
at large momentum transfer, yet mostly static in nature\cite{PhysRevB.82.195108}.}%
.
If these effects are important, a (QS){\it GW}+DMFT calculation that includes both non-local exchange {\it and} correlation effects must be performed.
We should however also note that {\it GW} itself misses the important non-local correlations that derive from spin-fluctuations.
Yet, interestingly, non-local spin fluctuations --e.g.\ those emerging in the vicinity of the N{\'e}el state of the 3D Hubbard model-- do not 
introduce a momentum differentiation to the dynamics of the self-energy but rather yield a modulation of
static contributions\cite{jmt_dga3d}. This reinforces the guidance that the separation \eref{separate}
may give to future electronic structure theories.

\subsection{Acknowledgments}
I acknowledge discussions with Gabi Kotliar and Mark van Schilfgaarde, the work with whom, Ref.~\cite{jmt_pnict}, 
provided the starting point for this paper.
Further, I thank Centre de Physique Th{\'e}orique, Ecole Polytechnique, for hospitality in the framework of a CNRS visiting position,
and my host Silke Biermann for years of fruitful discussions and collaborations.
I also acknowledge discussions with Ambroise van Roekeghem.
This work has been supported in part by the Research Unit FOR 1346 of the DFG (FWF Project ID I597-N16).

\section*{References}

%\bibliography{../../../refs,../../../refs_mine}

\begin{thebibliography}{10}
\expandafter\ifx\csname url\endcsname\relax
  \def\url#1{{\tt #1}}\fi
\expandafter\ifx\csname urlprefix\endcsname\relax\def\urlprefix{URL }\fi
\providecommand{\eprint}[2][]{\url{#2}}

\bibitem{bible}
Georges A, Kotliar G, Krauth W and Rozenberg M~J 1996 {\em Rev. Mod. Phys.\/}
  {\bf 68} 13

\bibitem{RevModPhys.78.865}
Kotliar G, Savrasov S~Y, Haule K, Oudovenko V~S, Parcollet O and Marianetti C~A
  2006 {\em Rev. Mod. Phys.\/} {\bf 78} 865--951

\bibitem{licht_katsnelson_kotliar}
Lichtenstein A~I, Katsnelson M~I and Kotliar G 2001 {\em Phys. Rev. Lett.\/}
  {\bf 87} 067205

\bibitem{PhysRevLett.86.5345}
Held K, Keller G, Eyert V, Vollhardt D and Anisimov V~I 2001 {\em Phys. Rev.
  Lett.\/} {\bf 86} 5345--5348

\bibitem{tomczak_v2o3_proc}
Tomczak J~M and Biermann S 2009 {\em J. Phys.: Condens. Matter\/} {\bf 21} 064209
  %preprint arXiv:0811.1098

\bibitem{me_psik}
Tomczak J~M and Biermann S 2009 {\em Phys. Status Solidi B (feature article)\/}
  {\bf 246} 1996 Scientific Highlight of the Month of the $\Psi_k$ Network, no.
  88, August 2008 %, arXiv:0907.1575

\bibitem{PhysRevLett.102.146402}
Kune\ifmmode~\check{s}\else \v{s}\fi{} J, Korotin D~M, Korotin M~A, Anisimov
  V~I and Werner P 2009 {\em Phys. Rev. Lett.\/} {\bf 102}(14) 146402
  %\urlprefix\url{http://link.aps.org/doi/10.1103/PhysRevLett.102.146402}

\bibitem{jmt_fesi}
Tomczak J~M, Haule K and Kotliar G 2012 {\em Proc. Natl. Acad. Sci. USA\/} {\bf
  109} 3243--3246 %preprint arXiv:1109.6561 (\textit{Preprint}
  %\eprint{http://www.pnas.org/content/109/9/3243.full.pdf+html})
  %\urlprefix\url{http://www.pnas.org/content/109/9/3243.abstract}

\bibitem{PhysRevLett.87.276404}
Held K, McMahan A~K and Scalettar R~T 2001 {\em Phys. Rev. Lett.\/} {\bf 87}
  276404

\bibitem{PhysRevLett.102.096401}
Pourovskii L~V, Delaney K~T, Van~de Walle C~G, Spaldin N~A and Georges A 2009
  {\em Phys. Rev. Lett.\/} {\bf 102}(9) 096401
  %\urlprefix\url{http://link.aps.org/doi/10.1103/PhysRevLett.102.096401}

\bibitem{jmt_cesf}
Tomczak J~M, Pourovskii L~V, Vaugier L, Georges A and Biermann S 2013 {\em
  Proc. Natl. Acad. Sci. USA\/} {\bf 110} 904--907 %preprint arXiv:1301.0630
  %(\textit{Preprint}
  %\eprint{http://www.pnas.org/content/110/3/904.full.pdf+html})
  %\urlprefix\url{http://www.pnas.org/content/110/3/904.abstract}

\bibitem{deltaPu}
Savrasov S~Y, Kotliar G and Abrahams E 2001 {\em Nature\/} {\bf 410} 793
  %\urlprefix\url{http://dx.doi.org/10.1038/35071035}

\bibitem{PhysRevB.84.195111}
Yin Q, Kutepov A, Haule K, Kotliar G, Savrasov S~Y and Pickett W~E 2011 {\em
  Phys. Rev. B\/} {\bf 84}(19) 195111
  %\urlprefix\url{http://link.aps.org/doi/10.1103/PhysRevB.84.195111}


\bibitem{RevModPhys.83.1589}
Stewart G~R 2011 {\em Rev. Mod. Phys.\/} {\bf 83}(4) 1589--1652
  %\urlprefix\url{http://link.aps.org/doi/10.1103/RevModPhys.83.1589}

\bibitem{1367-2630-11-2-025021}
Haule K and Kotliar G 2009 {\em New Journal of Physics\/} {\bf 11} 025021
  %\urlprefix\url{http://stacks.iop.org/1367-2630/11/i=2/a=025021}

\bibitem{PhysRevB.80.085101}
Aichhorn M, Pourovskii L, Vildosola V, Ferrero M, Parcollet O, Miyake T,
  Georges A and Biermann S 2009 {\em Phys. Rev. B\/} {\bf 80}(8) 085101
  %\urlprefix\url{http://link.aps.org/doi/10.1103/PhysRevB.80.085101}

\bibitem{PhysRevB.82.064504}
Aichhorn M, Biermann S, Miyake T, Georges A and Imada M 2010 {\em Phys. Rev.
  B\/} {\bf 82}(6) 064504
  %\urlprefix\url{http://link.aps.org/doi/10.1103/PhysRevB.82.064504}

\bibitem{Yin_pnictide}
{Yin} Z~P, {Haule} K and {Kotliar} G 2011 {\em Nat. Mat\/} {\bf 10} 932
  %preprint : arXiv1104.3454

\bibitem{PhysRevB.85.094505}
Ferber J, Foyevtsova K, Valent{\'i} R and Jeschke H~O 2012 {\em Phys. Rev. B\/}
  {\bf 85}(9) 094505
  %\urlprefix\url{http://link.aps.org/doi/10.1103/PhysRevB.85.094505}

\bibitem{werner_bfa}
Werner P, Casula M, Miyake T, Aryasetiawan F, Millis A~J and Biermann S 2012
  {\em Nat. Phys.\/} {\bf 8}(4) 331--337

\bibitem{annurev-conmatphys-020911-125045}
Georges A, Medici L~d and Mravlje J 2013 {\em Annual Review of Condensed Matter
  Physics\/} {\bf 4} 137--178 %(\textit{Preprint}
  %\eprint{http://dx.doi.org/10.1146/annurev-conmatphys-020911-125045})
  %\urlprefix\url{http://dx.doi.org/10.1146/annurev-conmatphys-020911-125045}

\bibitem{PhysRevB.81.220506}
Lee H, Zhang Y~Z, Jeschke H~O and Valent{\'i} R 2010 {\em Phys. Rev. B\/} {\bf
  81}(22) 220506
  %\urlprefix\url{http://link.aps.org/doi/10.1103/PhysRevB.81.220506}

\bibitem{PhysRevB.84.054529}
Aichhorn M, Pourovskii L and Georges A 2011 {\em Phys. Rev. B\/} {\bf 84}(5)
  054529 %\urlprefix\url{http://link.aps.org/doi/10.1103/PhysRevB.84.054529}

\bibitem{PhysRevLett.105.067002}
Borisenko S~V, Zabolotnyy V~B, Evtushinsky D~V, Kim T~K, Morozov I~V, Yaresko
  A~N, Kordyuk A~A, Behr G, Vasiliev A, Follath R and B\"uchner B 2010 {\em
  Phys. Rev. Lett.\/} {\bf 105}(6) 067002
  %\urlprefix\url{http://link.aps.org/doi/10.1103/PhysRevLett.105.067002}

\bibitem{PhysRevLett.110.167002}
Brouet V, Lin P~H, Texier Y, Bobroff J, Taleb-Ibrahimi A, Le~F\`evre P, Bertran
  F, Casula M, Werner P, Biermann S, Rullier-Albenque F, Forget A and Colson D
  2013 {\em Phys. Rev. Lett.\/} {\bf 110}(16) 167002
  %\urlprefix\url{http://link.aps.org/doi/10.1103/PhysRevLett.110.167002}

\bibitem{jmt_pnict}
Tomczak J~M, van Schilfgaarde M and Kotliar G 2012 {\em Phys. Rev. Lett.\/}
  {\bf 109}(23) 237010 %preprint arXiv:1209.2213
  %\urlprefix\url{http://link.aps.org/doi/10.1103/PhysRevLett.109.237010}

\bibitem{PhysRevLett.93.126406}
Faleev S~V, van Schilfgaarde M and Kotani T 2004 {\em Phys. Rev. Lett.\/} {\bf
  93} 126406

\bibitem{schilfgaarde:226402}
van Schilfgaarde M, Kotani T and Faleev S 2006 {\em Phys. Rev. Lett.\/} {\bf
  96} 226402 %\urlprefix\url{http://link.aps.org/abstract/PRL/v96/e226402}

\bibitem{RevModPhys.74.601}
Onida G, Reining L and Rubio A 2002 {\em Rev. Mod. Phys.\/} {\bf 74} 601--659

\bibitem{PhysRevLett.90.086402}
Biermann S, Aryasetiawan F and Georges A 2003 {\em Phys. Rev. Lett.\/} {\bf 90}
  086402

\bibitem{PhysRevLett.92.196402}
Sun P and Kotliar G 2004 {\em Phys. Rev. Lett.\/} {\bf 92} 196402

\bibitem{ayral_gwdmft}
Ayral T, Werner P and Biermann S 2012 {\em Phys. Rev. Lett.\/} {\bf 109}(22)
  226401 %\urlprefix\url{http://link.aps.org/doi/10.1103/PhysRevLett.109.226401}

\bibitem{PhysRevB.87.125149}
Ayral T, Biermann S and Werner P 2013 {\em Phys. Rev. B\/} {\bf 87}(12) 125149
  %\urlprefix\url{http://link.aps.org/doi/10.1103/PhysRevB.87.125149}

\bibitem{PhysRevLett.110.166401}
Hansmann P, Ayral T, Vaugier L, Werner P and Biermann S 2013 {\em Phys. Rev.
  Lett.\/} {\bf 110}(16) 166401
  %\urlprefix\url{http://link.aps.org/doi/10.1103/PhysRevLett.110.166401}

\bibitem{jmt_svo}
Tomczak J~M, Casula M, Miyake T, Aryasetiawan F and Biermann S 2012 {\em EPL\/}
  {\bf 100} 67001 %``EPL editor's choice'', preprint arXiv:1210.6580
  %\urlprefix\url{http://stacks.iop.org/0295-5075/100/i=6/a=67001}

\bibitem{jmt_svo_extended}
{Tomczak} J~M, {Casula} M, {Miyake} T and {Biermann} S 
2014 {\em Phys. Rev. B} {\bf 90} 165138
%{\em ArXiv  e-prints\/} 
%	(\textit{Preprint}
%  \eprint{1312.7546}), {\em Phys. Rev. B} (in press)

\bibitem{PhysRevB.88.165119}
Taranto C, Kaltak M, Parragh N, Sangiovanni G, Kresse G, Toschi A, Held K
 %{\em et al.} 
2013 {\em Phys. Rev. B} {\bf 88} 165119
%and Kaltak, M. and Parragh, N. and Sangiovanni, G. and Kresse, G. and Toschi, A. and Held, K.

\bibitem{PhysRevB.88.235110}
Sakuma R, Werner Ph. and Aryasetiawan F 2013 {\em Phys. Rev. B} {\bf 88}(23) 235110

\bibitem{paris_sex}
{van Roekeghem} A, {Ayral} T, {Tomczak} J~M, {Casula} M, {Xu} N, {Ding} H,
  {Ferrero} M, {Parcollet} O, {Jiang} H and {Biermann} S 2014 
	{\em Phys. Rev. Lett.} (in press),
	{\em ArXiv
  e-prints\/} 
	(\textit{Preprint} \eprint{1408.3136})

\bibitem{PhysRevB.83.054510}
Zhang Y {\em et al.}
%, Chen F, He C, Zhou B, Xie B~P, Fang C, Tsai W~F, Chen X~H, Hayashi H,  Jiang J, Iwasawa H, Shimada K, Namatame H, Taniguchi M, Hu J~P and Feng D~L
  2011 {\em Phys. Rev. B\/} {\bf 83}(5) 054510
  %\urlprefix\url{http://link.aps.org/doi/10.1103/PhysRevB.83.054510}

\bibitem{PhysRevB.76.165106}
Kotani T, van Schilfgaarde M and Faleev S~V 2007 {\em Phys. Rev. B\/} {\bf
  76}(16) 165106
  %\urlprefix\url{http://link.aps.org/doi/10.1103/PhysRevB.76.165106}

\bibitem{2014arXiv1406.0772I}
{Ismail-Beigi} S 2014 {\em ArXiv e-prints\/} (\textit{Preprint}
  \eprint{1406.0772})

\bibitem{PhysRevB.78.075106}
Sakuma R, Miyake T and Aryasetiawan F 2008 {\em Phys. Rev. B\/} {\bf 78}(7)
  075106 %\urlprefix\url{http://link.aps.org/doi/10.1103/PhysRevB.78.075106}

\bibitem{Sales2010304}
Sales B, McGuire M, Sefat A and Mandrus D 2010 {\em Physica C:
  Superconductivity\/} {\bf 470} 304 -- 308 ISSN 0921-4534
  %\urlprefix\url{http://www.sciencedirect.com/science/article/pii/S0921453410000651}

\bibitem{ferdi_down}
Aryasetiawan F, Tomczak J~M, Miyake T and Sakuma R 2009 {\em Phys. Rev.
  Lett.\/} {\bf 102} 176402 %preprint arXiv:0806.3373

\bibitem{PhysRevB.87.195144}
Hirayama M, Miyake T and Imada M 2013 {\em Phys. Rev. B\/} {\bf 87}(19) 195144
  %\urlprefix\url{http://link.aps.org/doi/10.1103/PhysRevB.87.195144}

\bibitem{PhysRevB.82.184511}
Yamasaki A {\em et al.}
% , Matsui Y, Imada S, Takase K, Azuma H, Muro T, Kato Y, Higashiya A,  Sekiyama A, Suga S, Yabashi M, Tamasaku K, Ishikawa T, Terashima K, Kobori H,  Sugimura A, Umeyama N, Sato H, Hara Y, Miyagawa N and Ikeda S~I 
	2010 {\em
  Phys. Rev. B\/} {\bf 82}(18) 184511
  %\urlprefix\url{http://link.aps.org/doi/10.1103/PhysRevB.82.184511}

\bibitem{PhysRevB.70.195104}
Aryasetiawan F, Imada M, Georges A, Kotliar G, Biermann S and Lichtenstein A~I
  2004 {\em Phys. Rev. B\/} {\bf 70} 195104

\bibitem{jmt_dga3d}
{Sch{\"a}fer} T, {Toschi} A and {Tomczak} J~M 2014 
{\em (submitted)\/}
%{\em ArXiv  e-prints\/} 
%	(\textit{Preprint}
%  \eprint{1312.7546})


\bibitem{PhysRevB.58.12684}
Schindlmayr A, Pollehn T~J and Godby R~W 1998 {\em Phys. Rev. B\/} {\bf 58}
  12684--12690

\bibitem{PhysRevB.82.195108}
Kang W and Hybertsen M~S 2010 {\em Phys. Rev. B\/} {\bf 82}(19) 195108
  %\urlprefix\url{http://link.aps.org/doi/10.1103/PhysRevB.82.195108}

\end{thebibliography}

\providecommand{\newblock}{}

\end{document}